\documentclass[a4paper,twoside,reqno]{bjp}

\usepackage{graphicx}
\usepackage{amssymb,amsmath,amscd,amsthm}
\usepackage{times}
\usepackage[bookmarks=false]{hyperref}
\hypersetup{%
    colorlinks=true,        
    linkcolor=blue,          
    citecolor=blue,         
    urlcolor=blue           
    }
\usepackage{geometry}
\allowdisplaybreaks

\usepackage[numbers,square,sort&compress]{natbib}
\usepackage{10be-shape-sdanca21}

\begin{document}

\title{\boldmath Symmetry and shape coexistence in $\isotope[10]{Be}$}

\runningheads{M.~A.~Caprio \textit{et al.}}{Symmetry and shape coexistence in $\isotope[10]{Be}$}

\begin{start}{%
\author{Mark A. Caprio}{1},
\author{Anna E. McCoy}{2},
\author{Patrick J. Fasano}{1},
\author{Tom\'{a}\v{s} Dytrych}{3,4}

\address{Department of Physics, University of Notre Dame, Notre Dame, Indiana 46556-5670, USA}{1}
\address{Institute for Nuclear Theory, University of Washington, Seattle, Washington 98195-1550, USA}{2}
\address{Nuclear Physics Institute of the Czech Academy of Sciences, 250\,68 \v{R}e\v{z}, Czech Republic}{3}
\address{Department of Physics and Astronomy, Baton Rouge, Louisiana State University, Louisiana 70803, USA}{4}

}

  \begin{Abstract}
    Within the low-lying spectrum of $\isotope[10]{Be}$, multiple rotational
    bands are found, with strikingly different moments of inertia.  A proposed
    interpretation has been that these bands variously represent triaxial
    rotation and prolate axially-deformed rotation.  The bands are
    well-reproduced in \textit{ab initio} no-core configuration interaction
    (NCCI) calculations.  We use the calculated wave functions to elucidate the
    nuclear shapes underlying these bands, by examining the Elliott $\grpsu{3}$
    symmetry content of these wave functions.  The \textit{ab initio} results
    support an interpretation in which the ground-state band, along with an
    accompanying $K=2$ side band, represent a triaxial rotor, arising from an
    $\grpsu{3}$ irreducible representation in the $0\hw$ space.  Then, the
    lowest excited $K=0$ band represents a prolate rotor, arising from an
    $\grpsu{3}$ irreducible representation in the $2\hw$ space.
  \end{Abstract}

\begin{KEY}
\textit{Ab initio} nuclear theory, no-core configuration interaction (NCCI),
no-core shell model (NCSM), shape coexistence, nuclear rotation, Elliott
$\grpsu{3}$ symmetry, Daejeon16 interaction
\end{KEY}
\end{start}


\section{Introduction}
\label{sec:intro}

The low-lying positive-parity spectrum of $\isotope[10]{Be}$ contains rotational
bands of decidedly different moments of
inertia~\cite{freer2007:cluster-structures,bohlen2007:10be-pickup}.  A ground
state $K=0$ band and excited $K=0$ band (at $\approx6.18\,\MeV$) are both
observed, through their $4^+$
members~\cite{freer2006:10be-resonant-molecule,bohlen2007:10be-pickup,suzuki2013:6he-alpha-10be-cluster},
and the moment of inertia for the excited band is higher by a factor of
$\approx3$.  Indeed, the difference of momenta of inertia is such that the
excited band hits the yrast line at $J=4$, and the $4^+_1$ state is a member of
the excited band.

Moreover, the ground-state band is accompanied by a proposed $K=2$ side band (at
$\approx5.96\,\MeV$), observed through its $3^+$
member~\cite{bohlen2007:10be-pickup}.  This band has a moment of inertia essentially
identical to that of the ground-state band.  Comparing the excitation energy of
the $2^+$ band head with that of the $2^+$ member of the ground-state band gives
$E(2^+_2)/E(2^+_1)\approx1.8$. Such a low ratio is suggestive of a triaxial
rotor with maximal triaxiality ($\gamma\approx
30^\circ$)~\cite{davydov1958:arm-intro,meyertervehn1975:triax-odda}.

The cluster interpretation provided by antisymmetrized molecular dynamics (AMD)
calculations~\cite{kanadaenyo1997:c-amd-pn-decoupling,kanadaenyo1999:10be-amd,suhara2010:amd-deformation}
is that $\isotope[10]{Be}$ consists of an $\alpha+\alpha$ dimer
($\isotope[8]{Be}$) plus two valence neutrons occupying molecular orbitals.  In
the ground-state and side bands, these two neutrons occupy equatorial orbitals
around the molecular axis ($\pi$ orbitals), and proton-neutron triaxiality
arises, in which an oblate distribution for the neutrons combines with the
prolate distribution of the protons (in the $\alpha+\alpha$ dimer) to form an
overall triaxial
shape~\cite{kanadaenyo1997:c-amd-pn-decoupling,suhara2010:amd-deformation}.  In
the excited $K=0$ band, the two valence neutrons instead occupy polar orbitals
along the molecular axis ($\sigma$ orbitals), which give rise to a large
axially-symmetric prolate deformation.

We turn here to \textit{ab initio} no-core configuration interaction
(NCCI)~\cite{navratil2000:12c-ab-initio,barrett2013:ncsm}, or no-core shell
model (NCSM), calculations for $\isotope[10]{Be}$ for further insight into the
nature of these states.  \textit{Ab initio} results can provide access to traditional collective
observables, such as electromagnetic transition strengths, which might otherwise
not be readily accessible to
experiment~\cite{caprio2013:berotor,maris2015:berotor2,caprio2015:berotor-ijmpe}.
Moreover, \textit{ab initio} calculated wave functions have been
found~\cite{dytrych2007:sp-ncsm-evidence,dytrych2013:su3ncsm} to have dominant
components with specific Elliott
$\grpsu{3}$~\cite{elliott1958:su3-part1,harvey1968:su3-shell} or
$\grpsptr$~\cite{rosensteel1977:sp6r-shell} symmetry.
\textit{Ab initio} calculations can thus give insight into the nature of collective states by exposing the symmetry
structure of calculated wave
functions~\cite{caprio2020:bebands,mccoy2020:spfamilies,zbikowski2021:beyond-elliott}.

\begin{figure}[tb]
\begin{center}
  \includegraphics[width=0.58\hsize]{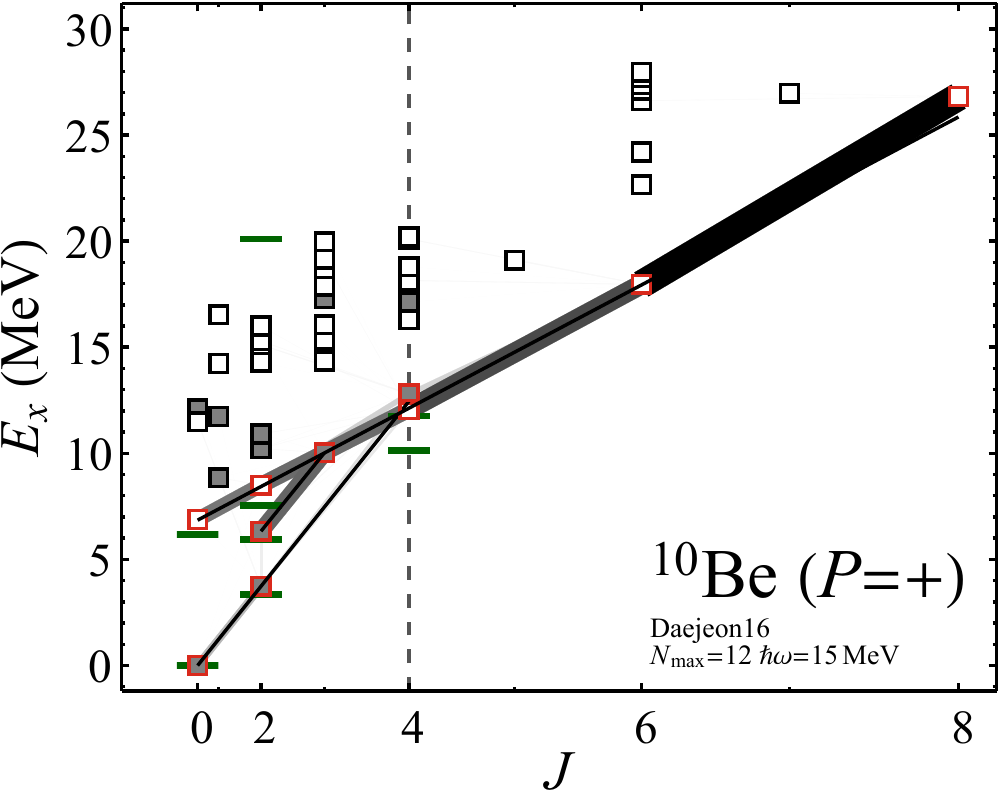}
  \hfill
\includegraphics[width=0.40\hsize]{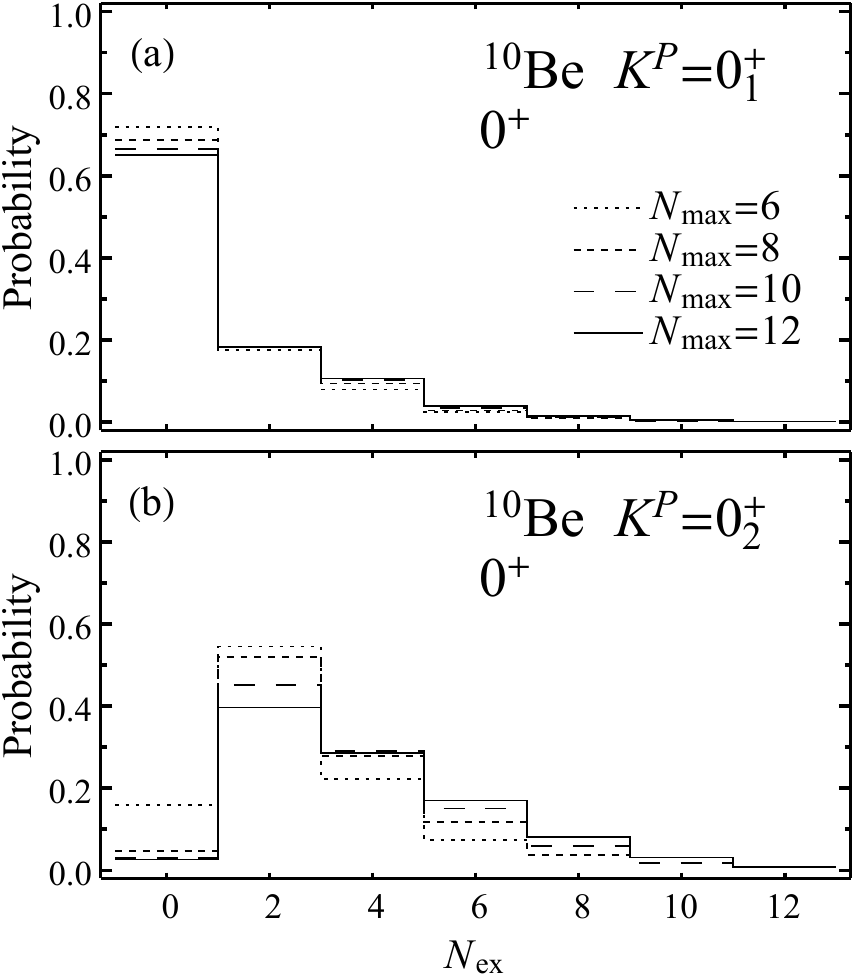}
\end{center}
\caption{(Left)~\textit{Ab initio} calculated energy spectrum of positive parity
  states for $\isotope[10]{Be}$, for $\Nmax=12$.  The $E2$ transition strengths
  (indicated by line widths) are shown for transitions from members of the
  ground-state band ($K=0$), side band ($K=2$), and excited $K=0$ band.  States
  are classified as $0\hw$ (filled squares) or $2\hw$ (open squares) by their
  $\Nex$ decomposition.  Experimental energies (thick horizontal lines),
  rotational energy fits (thin lines), and the maximal angular momentum for the
  $0\hw$ space (vertical dashed line) are shown for comparison.
  (Right)~Decompositions with respect to $\Nex$, for the calculated $0^+$ band
  head states of the (a)~ground-state band and (b)~long band.  Shown for
  $\Nmax=6$ to $12$ (dotted through solid curves).  Figure adapted
  from Ref.~\cite{caprio2019:bebands-sdanca19}.  }
\label{fig:levels-decompositions-10be}
\end{figure}

In a contribution~\cite{caprio2019:bebands-sdanca19} to the proceedings of the
previous workshop in this series, we explored the rotational features of the
\textit{ab initio} calculated spectrum of $\isotope[10]{Be}$.  The \textit{ab
  initio} NCCI calculated spectrum from Ref.~\cite{caprio2019:bebands-sdanca19},
obtained with the Daejeon16 interaction~\cite{shirokov2016:nn-daejeon16}, may be
seen in Figure~\ref{fig:levels-decompositions-10be}~(left).  The ground-state
band ($0^+$, $2^+$, $4^+$) and side band ($2^+$, $3^+$) are robustly reproduced,
terminating at the same angular momentum as in experiment.  An excited $K=0$
band with higher moment of inertia is also obtained, likewise in agreement with
experiment, but extending well past the last experimentally observed band
member, terminating rather at $8^+$.

Since these calculations are carried out in an oscillator basis, it was
straightforward to examine the contributions of configurations with different
numbers $\Nex$ of oscillator excitations, shown in
Figure~\ref{fig:levels-decompositions-10be}~(right).  For the ground-state and
side band members, such as the $0^+_1$ state
[Figure~\ref{fig:levels-decompositions-10be}~(right,a)], the dominant
contribution comes from $\Nex=0$ (or ``$0\hw$'') configurations, while, for members of the excited $K=0$ band, such as the $0^+_2$ state
[Figure~\ref{fig:levels-decompositions-10be}~(right,b)]], the dominant
  contribution comes from $\Nex=2$ (or ``$2\hw$'') configurations.

Here, we furthermore extract the Elliott $\grpsu{3}$ symmetry structure of the
same calculated wave functions considered in
Ref.~\cite{caprio2019:bebands-sdanca19}, to see what this information can
elucidate regarding the nature of the bands and their intrinsic nuclear shapes.
What does the $\grpsu{3}$ structure tell us about the deformation?  Do the
ground-state band and side band together represent the spectrum of a triaxial rotor?
After introducing the $\grpsu{3}$ irreducible representations (irreps)
arising in the shell-model space for $\isotope[10]{Be}$
(Section~\ref{sec:space}), we present results for the symmetry decompositions of
the rotational states in $\isotope[10]{Be}$ and discuss their implications
(Section~\ref{sec:results}).

\section{\boldmath $\grpsu{3}$ content of the many-body space for $\isotope[10]{Be}$}
\label{sec:space}

The nuclear many-body space reduces into irreducible representations (irreps) of
definite $\grpu{3}\times\grpsu{2}$ symmetry, described by the simultaneous
quantum numbers $\Nex(\lambda,\mu)S$.  Here $\grpu{3}$ is the symmetry group of
the harmonic oscillator.  Its irreps may be labeled by $\grpu{1}\times\grpsu{3}$
quantum numbers, where the $\grpu{1}$ group is generated by the harmonic
oscillator Hamiltonian, and yields the number of oscillator quanta as its
quantum number (we take $\Nex$ relative to the lowest Pauli-allowed oscillator
configuration), while the $\grpsu{3}$ group, generated by the orbital angular
momentum operators $L_{1,M}$ and Elliott's quadruople operators $\calQ_{2,M}$
(see, \textit{e.g.}, Appendix~A of Ref.~\cite{caprio2020:intrinsic}), gives the
Elliott $(\lambda,\mu)$ quantum numbers. These operators mutually commute with
the spin angular momentum operators $S_{1,M}$, which generates the usual spin
$\grpsu{2}$ group.

\begin{table}[tbp]  
  \caption{The $\grpu{3}\times\grpsu{2}$ irreps arising in the $0\hw$ space for
    $\isotope[10]{Be}$, ordered by increasing eigenvalue of the $\grpsu{3}$
    Casimir operator, together with selected irreps in the $2\hw$ space.  }
\label{tab:irreps-10be}
\small\smallskip
\tabcolsep=3.6pt
  \begin{tabular}{@{}lcr@{}}
\hline
\hline
\\[-8pt]
$\Nex(\lambda,\mu)$&$S$&$\tbracket{C_{\grpsu{3}}}$
\\
\hline
\\[-8pt]
    0(0,0)&0,1,2& 0\\ 
    0(1,1)&0,1,2& 6\\
    0(0,3)&1&     12\\
    0(3,0)&1&     12\\
    0(2,2)&0&     16\\
    \hline
    $\cdots$\\
    2(6,1)&0,1,2&   42.67\\
    2(8,0)&0&     58.67\\
\hline
\hline
  \end{tabular}
\end{table}
\begin{figure}[tbp]
\begin{center}
  \includegraphics[width=0.35\hsize]{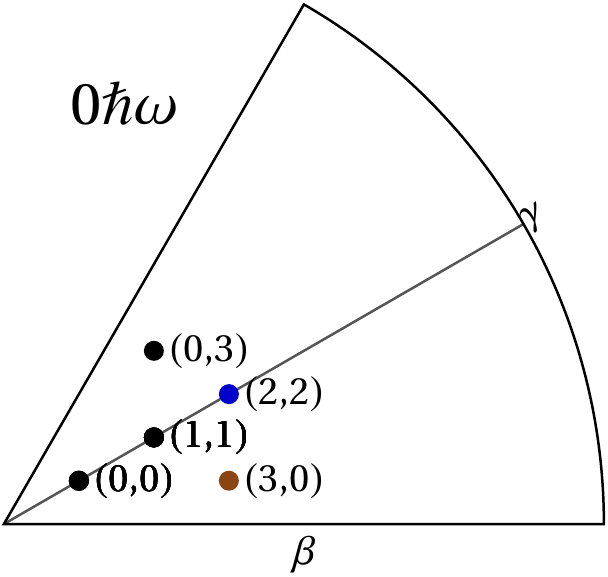}
  \hspace{0.05\hsize}
  \includegraphics[width=0.35\hsize]{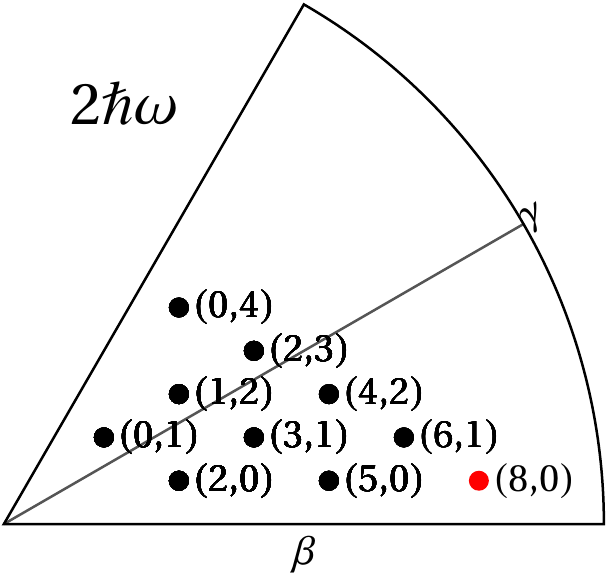}
\end{center}
\caption{Deformation $(\beta,\gamma)$ parameters for the $\grpu{3}$ irreps in
  the $0\hw$ (left) and $2\hw$ (right) spaces for $\isotope[10]{Be}$.  }
\label{fig:su3-shape-10be-g0}
\end{figure}
\begin{figure}[tbp]
\begin{center}
  \includegraphics[width=0.70\hsize]{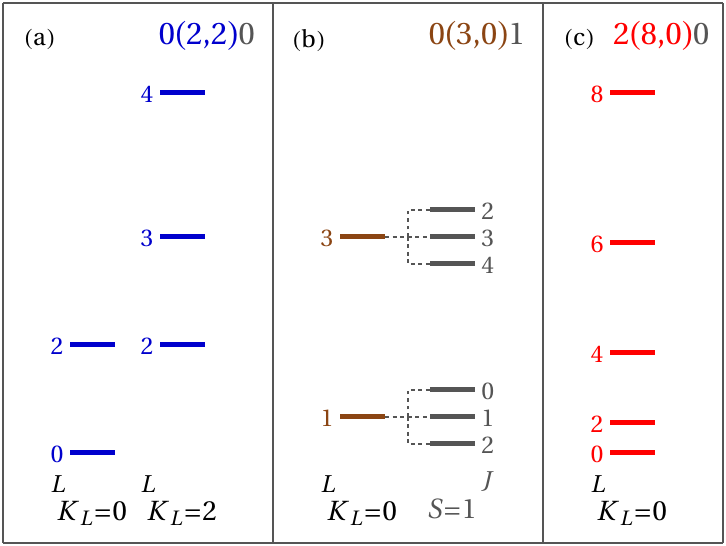}
\end{center}
\caption{Energy levels within the $\grpu{3}$ irreps relevant to the low-lying
  positive-parity spectrum of $\isotope[10]{Be}$: (a)~the leading irrep in the
  $0\hw$ space [$\Nex(\lambda,\mu)S=0(2,2)0$], (b)~the subleading irrep in the
  $0\hw$ space [identical spectra are obtained for $0(3,0)1$ and
    $0(0,3)1$], and (c)~the leading irrep of the $2\hw$ space
  [$2(8,0)0$].  In panel~(b), the orbital angular momenta $L$ arising within the
  $\grpu{3}$ irrep are shown at left, and the resultant total angular momenta $J$ from coupling with spin
  $S=1$ are shown at right.  }
\label{fig:su3-scheme-10be-g0}
\end{figure}

The $\grpu{3}\times\grpsu{2}$ irreps arising in the many-body space for a
nucleus may be deduced by general methods~\cite{draayer1989:un-u3-plethysm}
which are implemented in the code
\texttt{LSU3shell}~\cite{dytrych2016:su3ncsm-12c-efficacy}.  For
$\isotope[10]{Be}$, irreps from the $0\hw$ space (and selected irreps from the
$2\hw$ space) are listed in Table~\ref{tab:irreps-10be}.

The irrep within the many-body space having the largest eigenvalue $\tbracket{C_{\grpsu{3}}}$, of the $\grpsu{3}$ quadratic Casimir operator
\begin{math}
C_{\grpsu{3}} = \frac{1}{6}\left[\calQ\cdot\calQ +
  3{\vec{L}}\cdot{\vec{L}}\right],
\end{math}
is known as the \textit{leading irrep}.  This is the irrep which is most
strongly bound (energetically preferred) by Elliott's schematic
$-\calQ\cdot\calQ$ Hamiltonian~\cite{harvey1968:su3-shell}.  The concept of a
leading irrep was defined in the context of the traditional shell model, where
consideration was limited to the irreps in the $0\hw$ space.  However, we may
similarly identify a leading irrep in the $2\hw$ space, \textit{etc.}  For
$\isotope[10]{Be}$, the eigenvalues
\begin{math}
  \tbracket{C_{\grpsu{3}}}=\frac23(\lambda^2 + \lambda\mu + \mu^2 + 3\lambda + 3\mu)
\end{math}
are shown in Table~\ref{tab:irreps-10be} as well.

To each irrep of $\grpsu{3}$ is also associated a deformation $(\beta,\gamma)$, given by~\cite{castanos1988:su3-shape}
\begin{equation}
  \label{eq:beta-gamma}
    \beta^2 =\frac{4\pi}{5(A\overline{r^2})^2}\bigl(\lambda^2 + \lambda\mu + \mu^2 + 3\lambda + 3\mu + 3\bigr)
    \quad\quad
    \tan \gamma =\frac{\sqrt3 (\mu+1)}{2\lambda+\mu+3}.
\end{equation}
From this expression~(\ref{eq:beta-gamma}), we can see that the leading irrep
furthermore has the largest $\beta$ deformation.  The deformations
$(\beta,\gamma)$ for the $0\hw$ and $2\hw$ irreps of $\isotope[10]{Be}$ are
shown on a polar plot in Figure~\ref{fig:su3-shape-10be-g0}.

For $\isotope[10]{Be}$, the leading irrep in the $0\hw$ space (see
Table~\ref{tab:irreps-10be} and Figure~\ref{fig:su3-shape-10be-g0}) has quantum
numbers $\Nex(\lambda,\mu)S=0(2,2)0$, while the two subleading irreps,
degenerate with respect to the Casimir operator, have quantum numbers $0(3,0)1$
and $0(0,3)1$.  Irreps with quantum numbers $(\lambda,0)$ or $(0,\mu)$, such as
$(3,0)$ and $(0,3)$, have (nearly) prolate ($\gamma\approx0^\circ$) and (nearly)
oblate ($\gamma\approx60^\circ$) deformations, respectively, while those with
quantum numbers $(\lambda,\mu=\lambda)$, such as $(2,2)$, have a maximally
triaxial deformation ($\gamma=30^\circ$).  Then, the leading irrep within the
$2\hw$ space for $\isotope[10]{Be}$ has $\Nex(\lambda,\mu)S=2(8,0)0$.  This
irrep again is identified with a (nearly) prolate deformation, larger than for
the $0\hw$ irreps.

The levels within these irreps are obtained by considering the standard
$\grpsu{3}\rightarrow\grpso{3}$ branching relations, giving the allowed values
for the orbital angular momentum $L$~\cite{harvey1968:su3-shell}, which then
couple with the spin to give the total angular momentum $J$ ($L\times S\rightarrow
J$), as shown in Figure~\ref{fig:su3-scheme-10be-g0}.  The angular momentum
content of the leading $0(2,2)0$ irrep exactly matches that of the combined
ground-state and side bands in $\isotope[10]{Be}$ (the assignment of the $L=4$
terminating state to one band or the other, in the branching rule, is
arbitrary).  Then, an $(8,0)$ $\grpsu{3}$ irrep contains a single $K_L=0$ band,
extending through $L=8$.  The angular momentum content thus, notably, matches
that of the \textit{ab initio} calculated $2\hw$ excited $K=0$ band.

\section{\boldmath $\grpsu{3}$ nature of the low-lying states of $\isotope[10]{Be}$}
\label{sec:results}

\begin{figure}[p]
  \begin{center}
    \includegraphics[width=0.90\hsize]{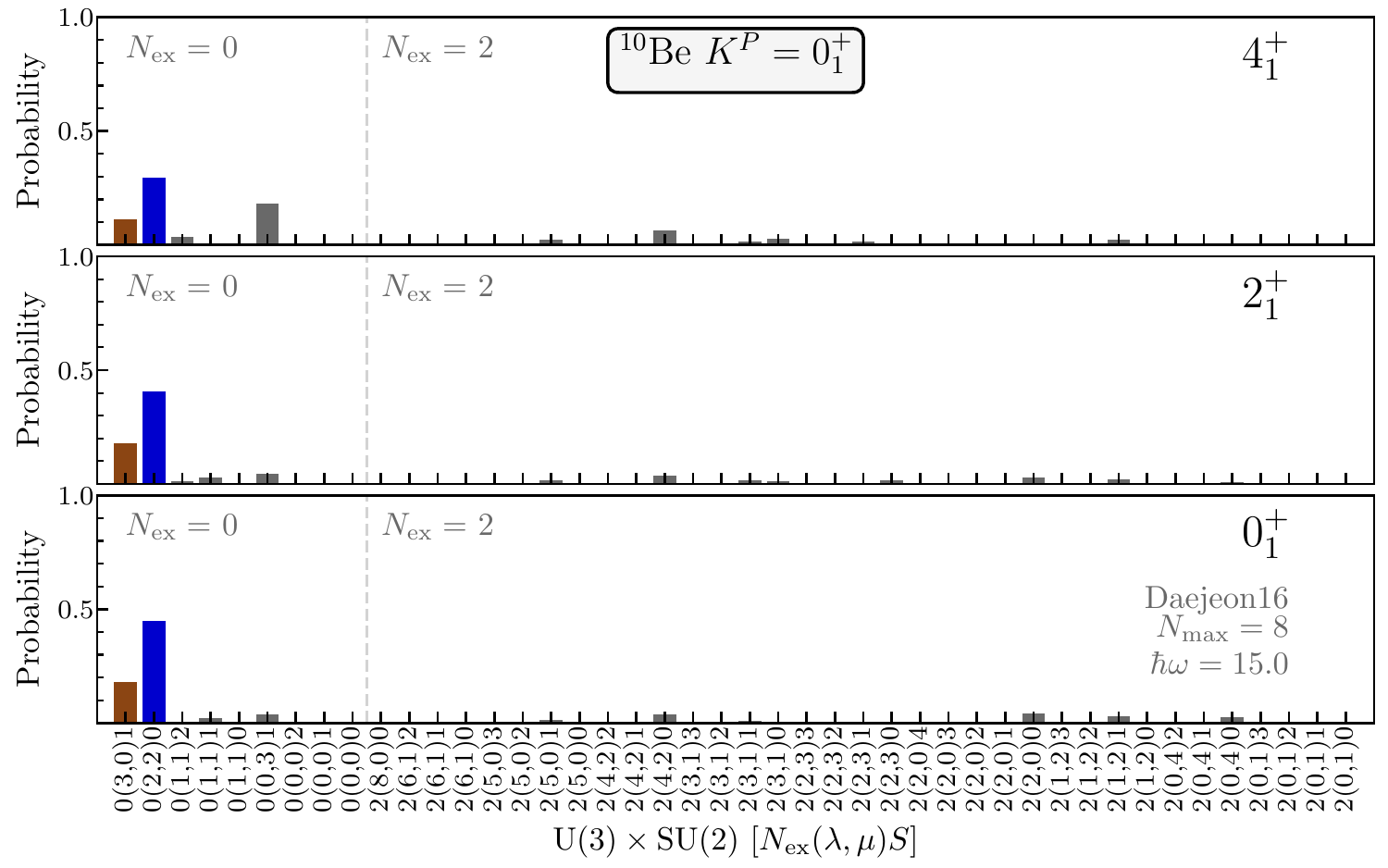}
    \\
    \includegraphics[width=0.90\hsize]{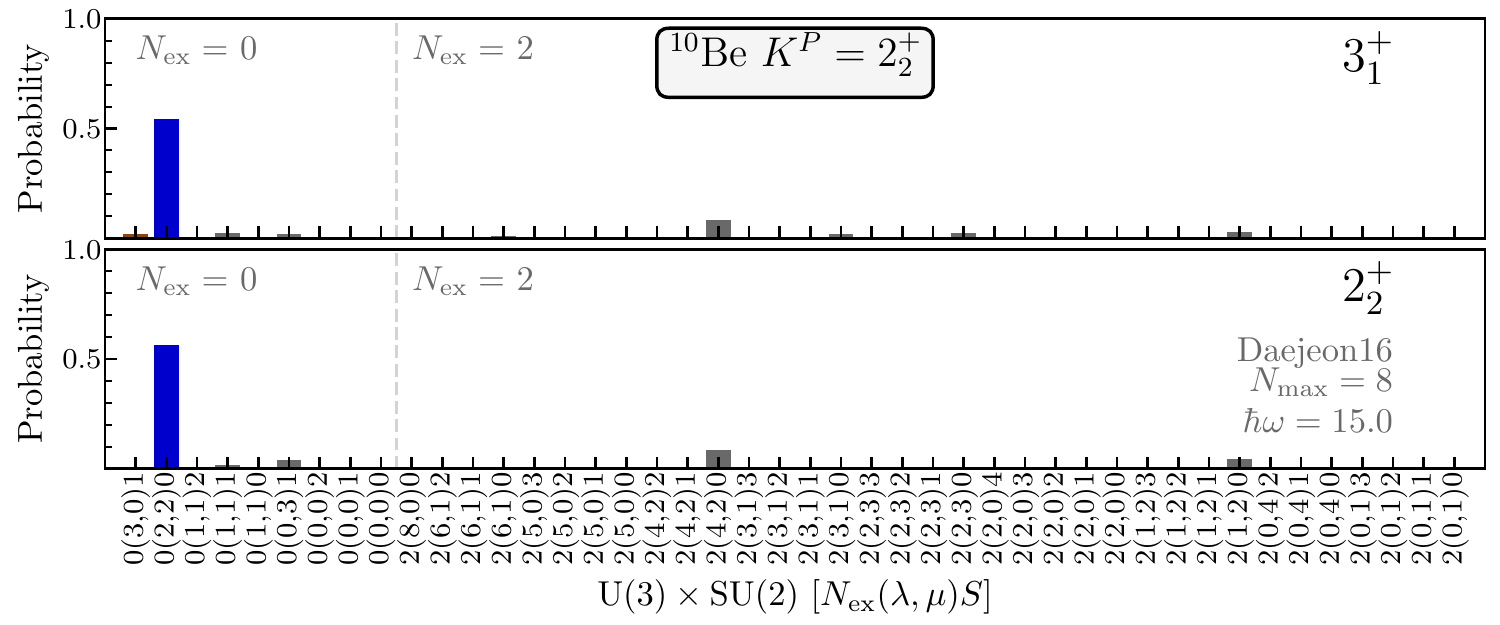}
    \\
    \includegraphics[width=0.90\hsize]{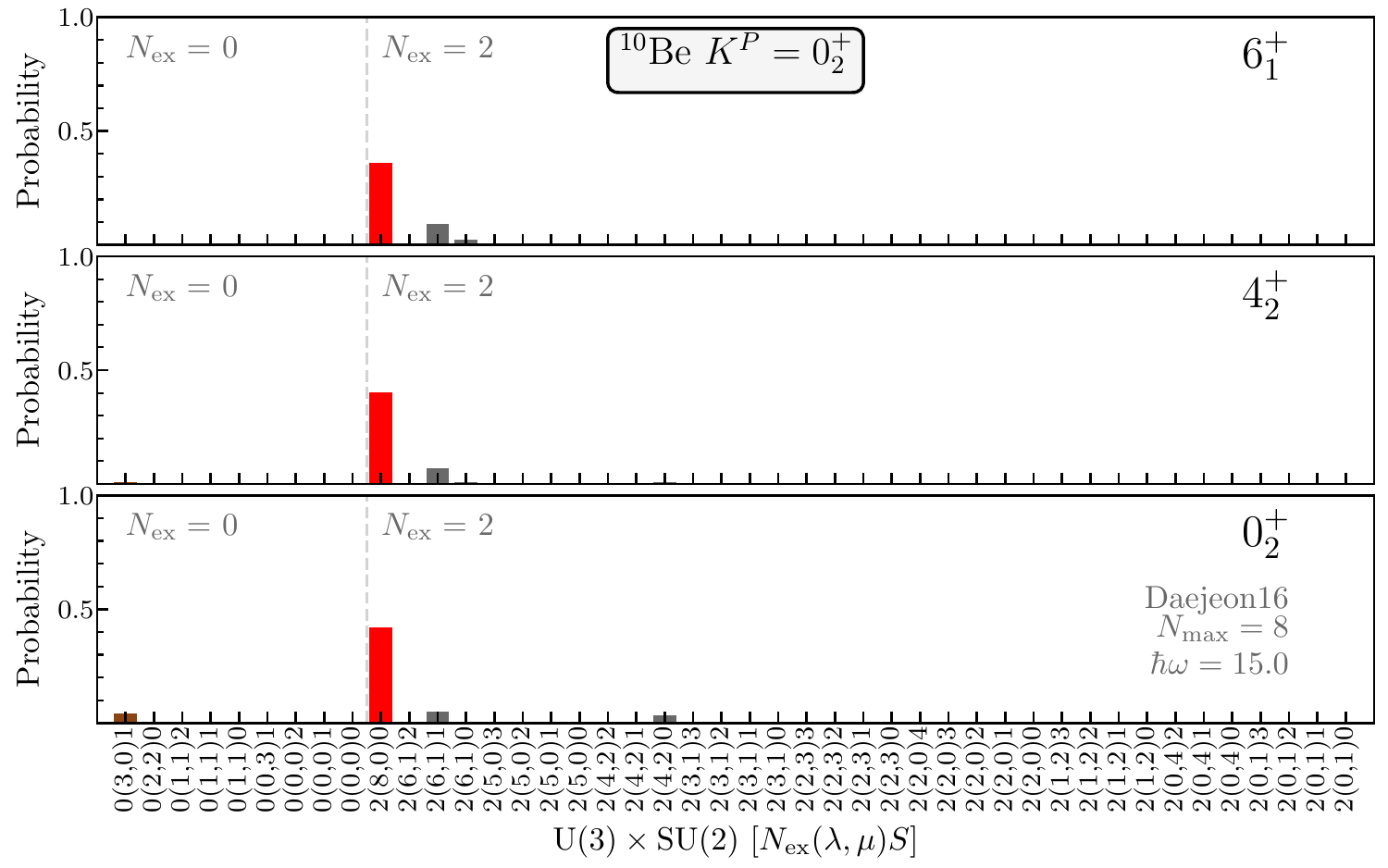}
\end{center}
\caption{Decompositions with respect to $\grpu{3}$ and spin quantum numbers for
  members of the ground-state band~(top), side band~(middle), and excited $K=0$
  band~(bottom) of $\isotope[10]{Be}$, calculated with $\Nmax=8$.
}

\label{fig:decompositions-10be}
\end{figure}

To decompose the \textit{ab initio} calculated wave function into contributions
from different $\grpu{3}\times\grpsu{2}$ symmetry subspaces, with quantum
numbers $\Nex(\lambda,\mu)S$, we make use of the ``Lanczos
trick''~\cite{whitehead1980:lanczos}.  In general, this technique provides the
decomposition of a wave function into its projections onto the eigenspaces of a
given Hermitian operator.  Thus, to obtain a decomposition with respect to $L$
or $S$, one decomposes the wave function into its projections onto eigenspaces
of the $\vec{L}^2$ or $\vec{S}^2$ operators~\cite{johnson2015:spin-orbit},
respectively.  For a decomposition with respect to $\grpsu{3}$ quantum numbers
$(\lambda,\mu)$, one decomposes the wave function into its projections onto
eigenspaces of the $\grpsu{3}$ quadruatic Casimir
operator~\cite{gueorguiev2000:fp-su3-breaking,herrera2017:cr-backbending-qds-su3-su4,zbikowski2021:beyond-elliott}.
(This may still leave some unresolved degeneracies if different irreps share the
same eigenvalue for $C_{\grpsu{3}}$.)

Here, to obtain simultaneous decompositions with respect to the full set of quantum numbers
$\Nex(\lambda,\mu)S$, we decompose the wave functions into
the eigenspaces of an operator
constructed as a linear combination
\begin{equation}
  \label{eq:decomposition-operator}
  C=a_{\grpu{1}}\Nex + a_{\grpsu{3}}C_{\grpsu{3}} + a_{\grpsu{2}}\vec{S}^2.
\end{equation}
The coefficients $a_i$ are chosen so as to avoid numerical near-degeneracies in
eigenvalues for this operator within the $\isotope[10]{Be}$ NCCI model
space.\footnote[1]{In fact, the decomposition which
  was carried out also included the proton spin $S_p$ and neutron spin $S_n$
  among the set of simultaneous quantum numbers (see, \textit{e.g.},
  Ref.~\cite{luo2013:su3cmf}), through inclusion of terms proportional to
  $\vec{S}_p^2$ and $\vec{S}_n^2$ in the operator~(\ref{eq:decomposition-operator}) with respect to which the decomposition is carried out.  Resolving $S_p$ and $S_n$
  help to lift certain degeneracies, \textit{e.g.}, although the
  $0(3,0)1$ and $0(0,3)1$ irreps for $\isotope[10]{Be}$ share the same
  eigenvalue of $C_{\grpsu{3}}$, they differ in whether their spin arises from
  the neutrons or protons, respectively.}  Both the initial diagonalization and
the subsequent Lanczos decomposition have been carried out
with the code
MFDn~\cite{aktulga2013:mfdn-scalability,shao2018:ncci-preconditioned}.

The resulting probability decompositions are shown for members of the various
bands in Figure~\ref{fig:decompositions-10be}. These decompositions are for wave
functions calculated with $\Nmax=8$.  For this lower $\Nmax$, note that the
calculated excited $K=0$ band still lies entirely above the ground-state band.
Thus the ground-state band member is $4^+_1$, and the excited $K=0$ band member
is $4^+_2$, opposite to the situation in
Figure~\ref{fig:levels-decompositions-10be} (and experiment).  In all these
states, the dominant contributions at low $\Nex$ are dressed by contributions at
higher $\Nex$, which account for about half of the norm [recall
  Figure~\ref{fig:levels-decompositions-10be}~(right)].

The ground-state band [Figure~\ref{fig:decompositions-10be} (top)] and side band
[Figure~\ref{fig:decompositions-10be} (middle)] members all have as their single
largest component the leading $0(2,2)0$ (triaxial) irrep of the $0\hw$ space.
The side band members are comparatively ``pure'', with nearly the entire
$\Nex=0$ contribution coming from $0(2,2)0$, while the ground-state band members
have significant admixtures of the subleading $0(3,0)1$ (prolate) and/or
$0(0,3)1$ (oblate) irreps.  Despite these secondary contributions, the overall
predominance of $0(2,2)0$ across these states provides a microscopic
justification for the interpretation of these bands in terms of a single
triaxial rotor spectrum.

Moreover, in $\isotope[10]{Be}$, the $0(2,2)0$ irrep is obtained by coupling a
proton $(2,0)$ (prolate) irrep with a neutron $(0,2)$ (oblate) irrep.  This
underlying structure for the $(2,2)$ irrep supports the interpretation of triaxiality in
$\isotope[10]{Be}$ as being proton-neutron in nature, arising from the
combination of prolate proton and oblate neutron distributions.

The decompositions for the excited $K=0$ band members
[Figure~\ref{fig:decompositions-10be} (bottom)], in turn, all have the
leading $2(8,0)0$ (prolate) irrep of the $2\hw$ space as by far their strongest
single contribution.  This irrep is obtained in the $\isotope[10]{Be}$ many-body
space by coupling a weakly prolate proton $(2,0)$ irrep with a more strongly
prolate neutron $(6,0)$ irrep, supporting the interpretation that the increased
deformation of the excited $K=0$ band is contributed primarily by the neutrons.

\section{Conclusion}
\label{sec:concl}

The rotational bands in the low-lying positive-parity spectrum of
$\isotope[10]{Be}$~--- as obtained from experiment, microscopic AMD
calculations, and \textit{ab initio} NCCI calculations~--- provide an example of
shape coexistence.  A modestly deformed triaxial rotor coexists with a more
strongly deformed prolate rotor.  In a shell-model interpretation, the states
composing the former structure are $0\hw$ states, while those of the latter
structure are $2\hw$ states, both dressed by higher oscillator excitations.

Decompositions of the NCCI wave functions, into contributions defined by
$\grpu{3}$ and spin quantum numbers, corroborate and elaborate upon this
picture.  The leading $\grpsu{3}$ irreps of the $0\hw$ and $2\hw$ spaces~---
$(2,2)$ and $(8,0)$, respectively~--- are found to define the band structure of
the low-lying spectrum and to provide the strongest contributions to the
calculated states.  The identification of the ground-state and side bands with
the $(2,2)$ irrep, in particular, supports their interpetation as representing 
triaxial rotation.


\section*{Acknowledgements}

Pieter Maris and James P.~Vary are gratefully acknowledged for valuable
discussions and for their collaboration on the calculations of
Ref.~\cite{caprio2019:bebands-sdanca19}.  This material is based upon work
supported by the U.S.~Department of Energy, Office of Science, under Award
Numbers DE-FG02-95ER40934 and DE-FG02-00ER41132. This research used
computational resources of the National Energy Research Scientific Computing Center
(NERSC), a U.S.~Department of Energy, Office of Science, user facility supported
under Contract~DE-AC02-05CH11231.

%


\end{document}